\documentstyle[epsfig]{aipproc}

\def\ensuremath#1{{\ifmmode #1 \else $#1$\fi}}
\newcommand{\eg}{ e.g.}
\newcommand{\msun}{\ensuremath{M_{\odot}}}
\newcommand{\nuc}[2]{\ensuremath{\mathrm {^{#2}#1}}}
\newcommand{\ttt}[1]{\ensuremath{\times 10^{#1}}}

\begin{document}
\title{The Impact of Nuclear Reaction Rate Uncertainties on Evolutionary
Studies of the Nova Outburst}

\author{W. Raphael Hix$^{\dag \S}$, Michael S. Smith$^{\S}$, Anthony
Mezzacappa$^{\S}$, Sumner Starrfield$^{*}$, Donald L. Smith$^{\ddag}$}
\address{$^{\dagger}$Department of Physics \& Astronomy, University of 
Tennessee, Knoxville, TN 37996-1200 \\
$^{\S}$Physics Division, Oak Ridge National Laboratory\thanks{Research at the Oak Ridge
National Laboratory is supported by the U.S. Department of Energy under
contract DE-AC05-96OR22464 with Lockheed Martin Energy Research Corp.} Oak
Ridge, TN 37831-6354\\
$^{*}$Department of Physics \& Astronomy, Arizona State University, Tempe, AZ 
85287-1504\\
$^{\ddag}$ Technology Development Division, Argonne National Laboratory,
Argonne, IL 60439}

\maketitle

\begin{abstract}
The observable consequences of a nova outburst depend sensitively on the
details of the thermonuclear runaway which initiates the outburst. One of the
more important sources of uncertainty is the nuclear reaction data used as
input for the evolutionary calculations. A recent paper by Starrfield, Truran,
Wiescher, \& Sparks \cite{STWS98} has demonstrated that changes in the reaction rate
library used within a nova simulation have significant effects, not just on the
production of individual isotopes (which can change by an order of magnitude),
but on global observables such as the peak luminosity and the amount of mass
ejected.  We will present preliminary results of systematic analyses of the
impact of reaction rate uncertainties on nova nucleosynthesis.  
\end{abstract}

\section{Nuclear Uncertainties in Novae}
Observations of nova outbursts have revealed an elemental composition 
that differs markedly from solar.  Theoretical studies indicate that these 
differences are caused by the combination of convection with explosive 
hydrogen burning which results in a unique nucleosynthesis that is rich in 
odd-numbered nuclei such as $\nuc{N}{15}$, $\nuc{O}{17}$ and $\nuc{C}{13}$.  
These nuclei are difficult to form in other astrophysical events.  Many of 
the proton-rich nuclei produced in nova outbursts are radioactive, offering 
the possibility of direct observation with $\gamma$-ray instruments.  
Potentially important $\gamma$-ray producers include \nuc{Al}{26}, 
\nuc{Na}{22}, \nuc{Be}{7} and \nuc{F}{18}.

The observable consequences of a nova outburst depend sensitively on the
details of the thermonuclear runaway which initiates the outburst. One of the
more important sources of uncertainty is the nuclear reaction data used as
input for the evolutionary calculations \cite{STWS98}. A number of features
conspire to magnify the effects of nuclear uncertainties on nova
nucleosynthesis. Many reactions of relevance to novae involve unstable
proton-rich nuclei, making experimental rate determinations difficult. For
hydrodynamic conditions typical of novae, many rates depend critically on the
properties of a few individual resonances, resulting in wide variation between
different rate determinations. Statistical model (Hauser-Feshbach)
calculations, which are employed with great success for a large number of
reactions \cite{RaTK97a}, are unreliable for rates dominated by individual
resonances. The similarity of the nuclear burning and convective timescales
results in nuclear burning in novae which is far from the steady state which
typifies quiescent burning.

\section{Monte Carlo Estimates of Uncertainties}

Though analysis of the impact of variations in the rates of a few individual
reactions has recently been performed using one-dimension hydrodynamic
models \cite{JoCH99}, analysis of the impact of the complete set of possible
reaction rate variations in such hydrodynamic models remains computationally
prohibitive. We therefore begin by examining in detail the nucleosynthesis of
individual zones, using hydrodynamic trajectories (temperature and density as a
function of time) drawn from nova outburst models. Such one zone,
post-processing nucleosynthesis simulations are a common means of estimating
nova nucleosynthesis (see \eg, \cite{HiTh82,WGTR86}). For this presentation we
are using a hydrodynamic trajectory for an inner zone of a $1.35 \msun$ ONeMg
WD which is similar to that described in \cite{PSTW95}.  These calculations
were performed using a nuclear network with 87 species, composed of elements
from n and H to S, including all isotopes between the proton drip line and
the most massive stable isotope.

\begin{figure}[ht]
  \centering
  \epsfig{file=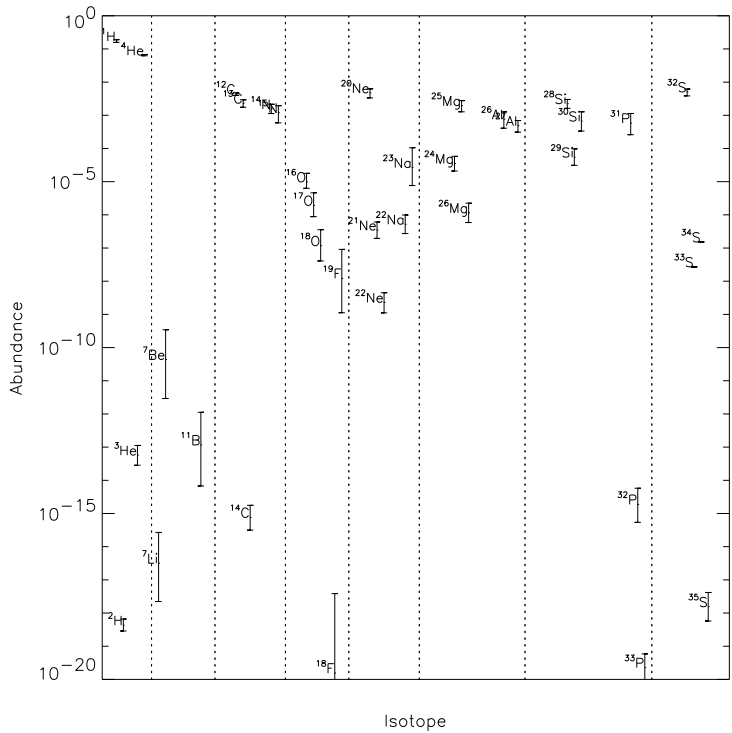,width=4.0in}
  \caption{Final Abundances (elapsed time = $5.2\ttt{5} \sec$ after peak).}
  \label{fig:mcabund}
\end{figure}

Figure~\ref{fig:mcabund} shows the abundances of each species at the end of the
simulation, $5.2\ttt{5} \sec$ after peak temperature. Because of the long time
which has elapsed, the unstable proton-rich nuclei have decayed, reducing their
abundances to less than $10^{-20}$.   To investigate the extent to which
nuclear reaction uncertainties translate into abundance differences, we use a
Monte Carlo technique which assigns to each reaction rate in the nuclear
network a random enhancement factor.  The error bars displayed in
Fig.~\ref{fig:mcabund} are the 90\% confidence intervals which result from
992 Monte Carlo iterations.  Monte Carlo methods have been employed with great
success in the analysis of Big Bang nucleosynthesis \cite{SmKM93}, but have not
previously been applied to other thermonuclear burning environments.

\begin{figure}[ht]
  \centering
  \epsfig{file=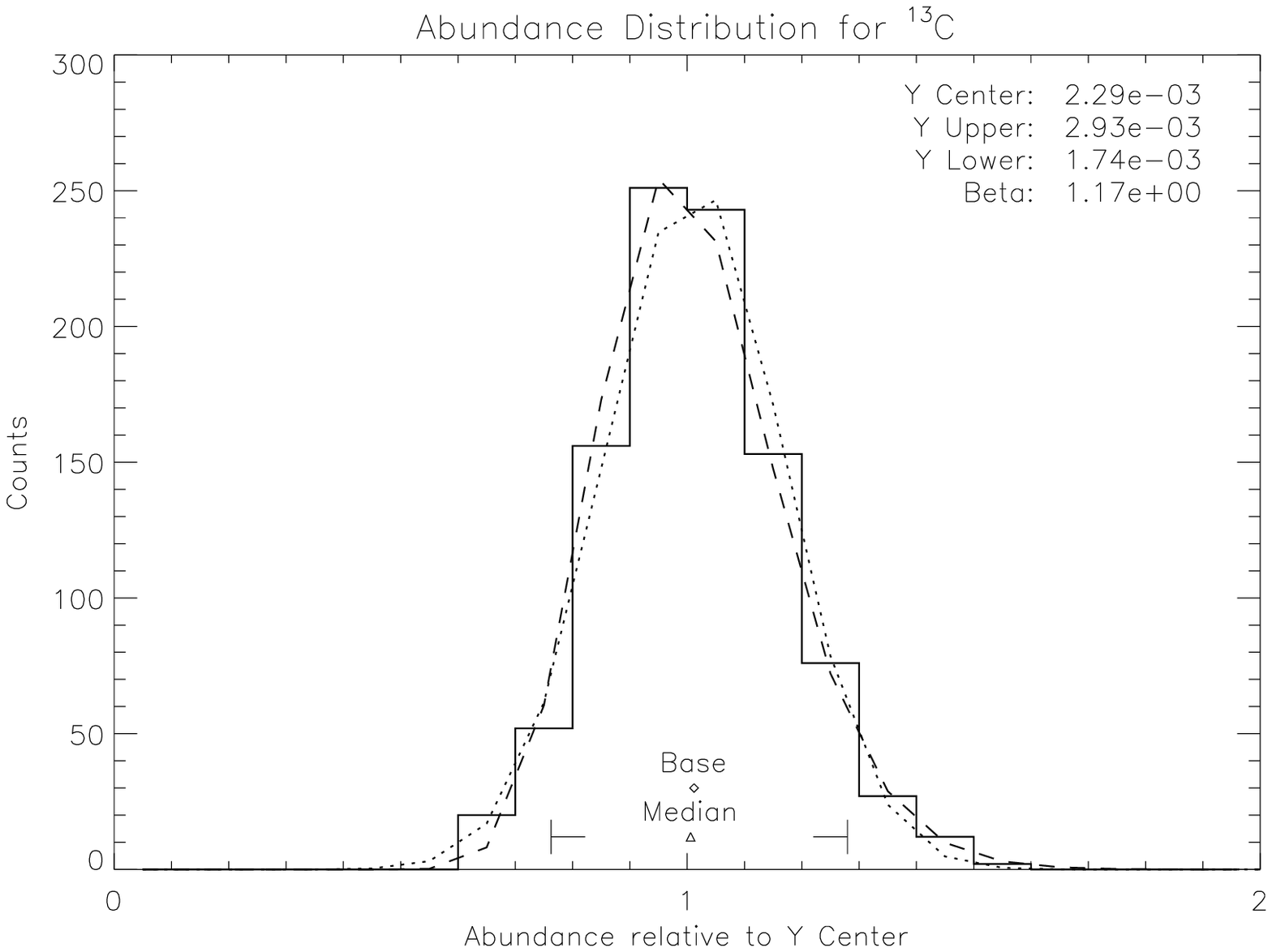,width=2.5in}
  \epsfig{file=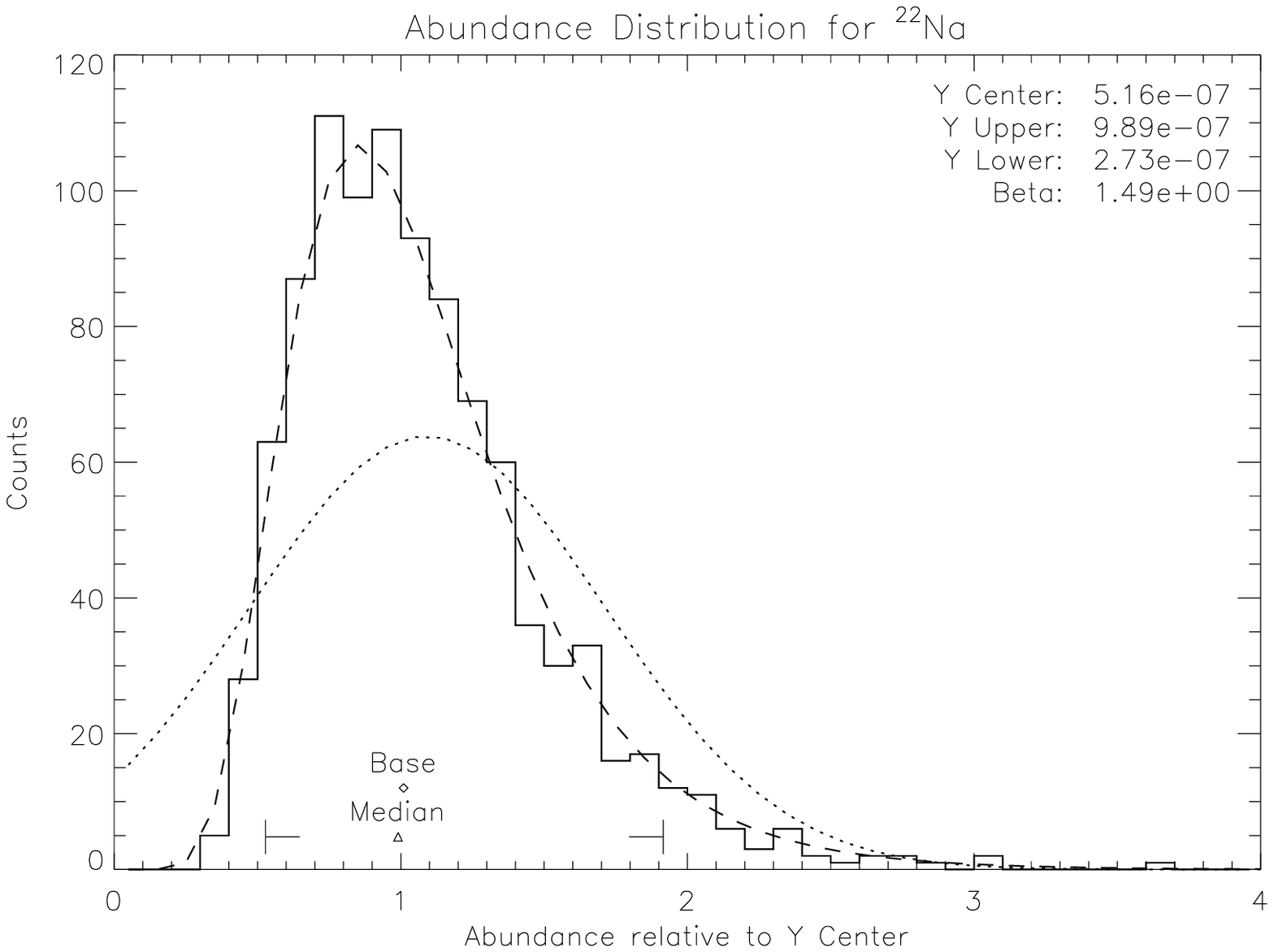,width=2.5in}
  \caption{The histograms showing deviations in abundance from the mean.   The
  dotted curves are normal distributions with the same (arithmetic) mean and
  standard deviation as the Monte Carlo distribution. The dashed curves are
  log-normal distributions with (logarithmic) mean and standard deviation from
  the Monte Carlo.  Y Center, Y Upper, and Y Lower correspond, respectively, to
  the central abundance and the upper and lower error bars from
  Fig.~\ref{fig:mcabund}.}
  \label{fig:hist}
\end{figure}

The reaction rate enhancement factors are distributed according to the
log-normal distribution, which is the correct uncertainty distribution for
quantities like reaction rates which are manifestly positive \cite{Smit91},
\begin{equation}
    p_{log-normal}(x) = \frac{1}{\sqrt{2} \, \pi \beta x}
    \exp \left(- \frac{(\ln x - \alpha)^2}{2 \beta^2} \right) \ ,
    \label{eq:lognorm}
\end{equation}
where $\alpha$ and $\beta$ are the (logarithmic) mean and standard deviation.
For small uncertainties $(<20\%)$, the difference between the log-normal
distribution and the normal (Gaussian) distribution is small. However, for
uncertainties of larger sizes such as those encountered in this problem, the
difference is important.  For this preliminary analysis, we have chosen to
assign uncertainties of $\sim 50\% \ (\beta=\ln(1.5))$ both to rates calculated by
Hauser-Feshbach methods and also to rates whose measurement require
radioactive ion beams.  For all other rates we assign $\beta=\ln(1.2)$. 
Figure~\ref{fig:hist} plots the resulting abundance distributions for two
representative nuclei.  Fig.~\ref{fig:hist} also demonstrates the differences
between normal and log-normal distributions for widths of these sizes. These
are very conservative uncertainties; relatively few  reactions, especially
among unstable nuclei, have measurement uncertainties this small.

\section{Results}

As evidenced by the error bars in Fig.~\ref{fig:mcabund}, the impact of even
our conservatively chosen variations in reaction  rates on the nucleosynthesis
is large.  While broader conclusions will require  analysis of additional
hydrodynamic trajectories, a number of interesting points can be made from the
analysis of this single trajectory. The impact on the rate of energy
production is small.  At the 90\% confidence level, variations in the amount
of hydrogen consumed represent $\sim 10\%$  variation in the thermonuclear
energy released.   For the most abundant metals (those which represent more
than 1\% of the mass), 2$\sigma$ variations by factors of 1.1 to 1.4 are
common, with some of these nuclei showing 2$\sigma$ variations as large as a
factor of 2, for example, \nuc{N}{15} ($2.1 \times$) and \nuc{Si}{30} ($2.3
\times$). For the $\gamma$-ray source nuclei \nuc{Na}{22} and \nuc{Al}{26},
the 90\% confidence interval includes variations of nearly a factor of 2,
representing  almost a factor of 4 difference in the distance from which novae
may be observed by $\gamma$-ray telescopes.  For \nuc{Be}{7}, the 90\%
confidence interval spans more than 2 orders of magnitude.

Such large uncertainties in the nucleosynthesis, resulting from poorly known
nuclear reaction rates, constrain our ability to make detailed comparisons
between theoretical models for the nova outburst and astrophysical observations to
a degree which is often ignored. Improved knowledge of these uncertain rates,
both experimental and theoretical, is necessary to provide tight constraints on
the nova outburst from its nucleosynthetic products. 


\end{document}